\documentclass[a4paper,12pt,twoside]{cpc-hepnp}

\usepackage{bm}
\usepackage{graphics}
\usepackage{graphicx}
\usepackage{caption}
\usepackage{epsfig}
\usepackage{amssymb}
\usepackage{amsmath}
\usepackage{amssymb}
\usepackage{slashed}
\usepackage{multirow}
\usepackage{epsfig}
\usepackage[tight,hang]{subfigure}
\usepackage{tikz-feynman,contour}
\usepackage{verbatim}

\catcode`\ö=\active
\defö{\"{o}}

\catcode`\ý=\active
\defý{\i}

\newcommand{\be}{\begin{equation}}
	\newcommand{\ee}{\end{equation}}
\newcommand{\bea}{\begin{eqnarray}}
	\newcommand{\eea}{\end{eqnarray}}
\newcommand{\bdm}{\begin{displaymath}}
	\newcommand{\edm}{\end{displaymath}}
\newcommand{\baa}{\begin{array}}
	\newcommand{\eaa}{\end{array}}
\newcommand{\ds}{\displaystyle}
\newcommand{\ec}{\slashed{\epsilon}}
\newcommand{\kc}{\slashed{k}}
\newcommand{\pc}{\slashed{p}}
\newcommand{\qc}{\slashed{q}}

\newcommand{\ba}{\begin{eqnarray}}
	\newcommand{\ea}{\end{eqnarray}}
\newcommand{\mL}{\mathcal{L}}
\newcommand{\mT}{\mathcal{T}}
\newcommand{\mM}{\mathcal{M}}
\newcommand{\mQ}{\mathcal{Q}}
\newcommand{\cu}{\mathpzc{u}}
\newcommand{\cv}{\mathpzc{v}}
\DeclareMathAlphabet{\mathpzc}{OT1}{pzc}{m}{it}

\begin{document}

		\fancyhead[c]{\small Chinese Physics C~~~Vol. XX, No. X (20XX)
			010201} \fancyfoot[C]{\small 010201-\thepage}
		
		\footnotetext[0]{Received XX Feb 2022}
		
		\title{Stellar Energy Loss Rates from Photoneutrino Process in Minimal Extension Standard Model}
		
		\author{%
			C. Ayd{\i}n\email{coskun@ktu.edu.tr}
		}
		\maketitle
		
		\address{%
			Department of Physics, Karadeniz Technical University, 61080, Trabzon, Turkey
		}

		\begin{abstract}
			Using the minimal extension  of Standard Model taking into account the charge radius and the anapole moments of neutrino, we have derived analytic expressions for the stellar energy loss rates due to the production of the neutrino pair process $e^- + \gamma \rightarrow e^- + \nu_e +\overline{\nu_e}$ in  a hot plasma for three limiting regimes (nondegenerate, intermediate and degenerate electrons) of temperature, the electron's chemical potential and plasma's energy.  Obtained results show that there is approximately extra $10\%$ contribution by using the considered calculations.
		\end{abstract}
		
		
			
\section{Introduction}

Energy loss due to neutrino emission is an important process astrophysical problems. Neutrinos are produced in stellar interiors, because of their interacting with stellar matter weakly, they can easily absorb energy that would otherwise take much longer to be transported to the surface by radiation or convection. 
The resulting energy sink in the center of the star can dictate the star's rate of nuclear burning, structure and evolution, and  ultimately how its life ends.
In this cases, if any process produce neutrino pairs, it is an extremely important energy-loss mechanism for stars in certain density and temperature ranges. 
The photoneutrino process were first studied by Ritus \cite{ref1} and  by Chiu and Stabler \cite{ref2} for non-degenerated as well as degenerate electron in Fermi $V-A$ theory. Beaudet, Betrosian and Salpeter\cite{ref3} provided analytic approximations for the photo-neutrino process in the same model.  Dicus \cite{ref4} recalculated this process in the framework of Standard Model(SM). He gave global correction factors to these rates to include neutral current effects. Schinder et al.\cite{ref5} numerically recalculated the emission rates in SM and  found good agreement with The BPS formulae together with the temperature range  $10^{10} -10^{11}K$. In 1989, Itoh  et al.\cite{ref6} whose improved  on their previous work \cite{ref7}  provided an alternate set of approximation formulae on their previous work. All above studies, the photon is taken ordinary photon. As regards ordinary photon this process is supposed to play an important role for the low  density $\rho/\mu_e =  10^5 gr/cm^3$ and comparatively low temperature  $T = 4\cdot10^8K$. These calculation has been done for about sixty years ago by many authors in $V-A$ model\cite{ref1,ref2, ref3}, SM \cite{ref4, ref5, ref6, ref7, ref8,ref9,ref10,ref11,ref12} and magnetic model\cite{ref9a}. Dutta  et al.\cite{ref10,ref11} considered instead of ordinary photon the massive photon (plasmon) and the angular dependence of the emitted neutrinos for this process in hot and dense matter. In prior their study didn't consider the energy and angular dependencies of the  neutrinos in the calculations of  the total energy lost rates, because they  used Lenard's formula \cite{ref13}. They presented numerical results for  widely varying conditions of temperatures and density.

In recent studies, we have investigated the production neutrino pair process  of the energy loss\cite{ref14}  and energy deposition rate for neutrino pair process \cite{ref14a}in the minimal extension standard model.Also,the  photo-neutrino, plasmon and bremstrahlung processes are the dominant cause of the stellar energy loss rates in different regions present within the  density-temperature plane.These  new calculations could contribute to a better understanding of the neutrino physics, and looking for new physics beyond standard model(SM). 

In this study, we have investigated the  energy loss rate of photo-neutrino pair annihilation process  $e^- + \gamma \rightarrow e^- + \nu_e +\overline{\nu}_e$ in the extension SM includes neutrino electromagnetic form factors (especially  charge radius and anapole moment) effect on neutrino  photon interaction in to account the $\gamma \nu \overline{\nu}$ vertex for the Dirac neutrinos \cite{ref14,ref14a,ref14b,ref15,ref15a,ref15b,ref15c,ref16,ref17a,ref17b,ref18,ref19,ref20,ref21,ref22,ref27,ref28} in Section II. These effects not considered previously for this process in literature. In Section III, we present our the numerical results. This section also contain discussion and our conclusion.

\section{Calculation}

The Feynman diagrams for the process $e^-(p) + \gamma (k) \rightarrow e^-(p') + \nu_e(q_1) +\overline{\nu}_e(q_2)$ are shown in Figure~1.The symbols in the parentheses are the momenta of the particles.

After Fierz transformation, the matrix element of the process can be written in SM  (Figure~1 (a)-(d)) in the low energy limit as

\begin{figure}[!h]
	\centering
	\subfigure[]{\begin{tikzpicture}[scale=0.95]
			\node at (-0.3,1.0) {\color{black}{}};
			\node at (3.3,0.7) {\color{black} {}	};
			\node at (-0.1,-0.85) {\color{black}{}	};
			\node at (3.3,-0.7) {\color{black}{}	};
			\begin{feynman}
				\vertex (a) at (-1.,1.2) ;
				\vertex (d) at ( 0, 0) ;
				\vertex (e) at (-1.,-1.2) ;
				\vertex (f) at (1.6,0) ;
				\vertex (g) at (2.6,-0.8) ;
				\vertex (h) at (3.,1.) ;
				\vertex (i) at (3.5,-0.5) ;
				\vertex (j) at (3.5,-1.3) ;
				\diagram* {
					(a) -- [fermion,edge label=$e^-$] (d) ,
					(e)--[photon,edge label=$\gamma$] (d), 
					(d)--[fermion, edge label=$e^-$] (f) -- [scalar,edge label=W] (g),
					(f)--[fermion,edge label=$\nu$] (h);
					(g)--[fermion,edge label=$e^-$] (i);
					(j)--[fermion,edge label=$\overline\nu$] (g)};
			\end{feynman}
	\end{tikzpicture}}
	\subfigure[]{
		\begin{tikzpicture}[scale=0.95]
			\node at (1,1.3) {\color{black}{}};
			\node at (2.9,1.3) {\color{black} {}	};
			\node at (-0.1,-0.85) {\color{black}{}	};
			\node at (3.3,-0.7) {\color{black}{}	};
			\begin{feynman}
				\vertex (a) at (-0.8,0.3) ;
				\vertex (d) at ( 0, 0) ;
				\vertex (e) at (-0.8,-2.) ;
				\vertex (f) at (2,0) ;
				\vertex (g) at (.8,-0.7) ;
				\vertex (h) at (1,0.4) ;
				\vertex (i) at (2,-0.3) ;
				\vertex (j) at (0.99,-1.5) ;
				\vertex (k) at (2,-1.6) ;
				\diagram* {
					(e) -- [photon,edge label=$\gamma$] (j) ,
					(a)--[fermion,edge label=$e^-$] (d), 
					(d)-- [scalar,edge label=W] (g),
					(d)--[fermion,edge label=$\nu$] (h);
					(i)--[fermion,edge label=$\overline\nu$] (g);
					(j)--[fermion,edge label=$e^-$] (k);

					(g)--[fermion,edge label=$e^-$] (j)};
			\end{feynman}
		\end{tikzpicture}	
	}
	\subfigure[]{
		\begin{tikzpicture}[scale=0.95]
			\node at (-0.3,1.0) {\color{black}{}};
			\node at (3.3,0.7) {\color{black} {}	};
			\node at (-0.1,-0.85) {\color{black}{}	};
			\node at (3.3,-0.7) {\color{black}{}	};
			\begin{feynman}
				\vertex (a) at (-0.8,1.2) ;
				\vertex (d) at ( 0, 0) ;
				\vertex (e) at (-0.8,-1.2) ;
				\vertex (f) at (2,0) ;
				\vertex (g) at (2.3,-0.8) ;
				\vertex (h) at (3.,1.) ;
				\vertex (i) at (3.5,-0.5) ;
				\vertex (j) at (3.5,-1.3) ;
				\diagram* {
					(e) -- [photon,edge label=$\gamma$] (d) ,
					(a)--[fermion,edge label=$e^-$] (d), 
					(d)--[fermion, edge label=$e^-$] (f) -- [scalar,edge label=Z] (g),
					(f)--[fermion,edge label=$e^-$] (h);
					(g)--[fermion,edge label=$\nu$] (i);
					(j)--[fermion,edge label=$\overline\nu$] (g)};
			\end{feynman}
		\end{tikzpicture}	
	} \\
	\subfigure[]{
		\begin{tikzpicture}[scale=0.95]
			\node at (-0.3,1.0) {\color{black}{}};
			\node at (3.3,0.7) {\color{black} {}	};
			\node at (-0.1,-0.85) {\color{black}{}	};
			\node at (3.3,-0.7) {\color{black}{}	};
			\begin{feynman}
				\vertex (a) at (-0.8,1.8) ;
				\vertex (d) at ( 0, 0) ;
				\vertex (e) at (-0.8,-0.8) ;
				\vertex (f) at (2,0) ;
				\vertex (g) at (2.2,0) ;
				\vertex (h) at (0,1.2) ;
				\vertex (i) at (3.,1.8) ;
				\vertex (j) at (3.0,.8) ;
				\vertex (k) at (1.6,1.2) ;
				\diagram* {
					(a) -- [fermion,edge label=$e^-$] (h) ,
					(e)--[photon,edge label=$\gamma$] (d), 
					(d)-- [fermion,edge label=$e^-$] (g),
					(h)--[fermion,edge label=$e^-$] (d);
					(k)--[fermion,edge label=$\nu$] (i);
					(h)--[scalar,edge label=Z] (k);

					(j)--[fermion,edge label=$\overline\nu$] (k)};
			\end{feynman}
		\end{tikzpicture}	
	}
	\subfigure[]{
		\begin{tikzpicture}[scale=0.95]
			\node at (-0.3,1.0) {\color{black}{}};
			\node at (3.3,0.7) {\color{black} {}	};
			\node at (-0.1,-0.85) {\color{black}{}	};
			\node at (3.3,-0.7) {\color{black}{}	};
			\draw [black,fill=black](2.5,-0.8) ellipse (.1cm and .18cm);
			\begin{feynman}
				\vertex (a) at (-0.8,1.2) ;
				\vertex (d) at ( 0, 0) ;
				\vertex (e) at (-0.8,-1.2) ;
				\vertex (f) at (1.6,0) ;
				\vertex (g) at (2.6,-0.8) ;
				\vertex (h) at (3.,1.2) ;
				\vertex (i) at (3.5,-0.5) ;
				\vertex (j) at (3.5,-1.3) ;
				\diagram* {
					(a) -- [fermion,edge label=$e^-$] (d) ,
					(e)--[photon,edge label=$\gamma$] (d), 
					(d)--[fermion, edge label=$e^-$] (f) -- [photon,edge label=$\gamma$] (g),
					(f)--[fermion,edge label=$e^-$] (h);
					(g)--[fermion,edge label=$\nu$] (i);
					(j)--[fermion,edge label=$\overline\nu$] (g)};
			\end{feynman}
		\end{tikzpicture}	
	}
	\subfigure[]{
		\begin{tikzpicture}[scale=0.95]
			\node at (-0.3,1.0) {\color{black}{}};
			\node at (3.3,0.7) {\color{black} {}	};
			\node at (-0.1,-0.85) {\color{black}{}	};
			\node at (3.3,-0.7) {\color{black}{}	};
			\draw [black,fill=black](1.6,1.5) ellipse (.1cm and .18cm);
			\begin{feynman}
				\vertex (a) at (-0.8,1.8) ;
				\vertex (d) at ( 0, 0) ;
				\vertex (e) at (-0.8,-1.0) ;
				\vertex (f) at (2,0) ;
				\vertex (g) at (2.2,0) ;
				\vertex (h) at (0,1.2) ;
				\vertex (i) at (3.0,2) ;
				\vertex (j) at (3.0,1) ;
				\vertex (k) at (1.6,1.5) ;
				\diagram* {
					(a) -- [fermion,edge label=$e^-$] (h) ,
					(e)--[photon,edge label=$\gamma$] (d), 
					(d)-- [fermion,edge label=$e^-$] (g),
					(h)--[fermion,edge label=$e^-$] (d);
					(k)--[fermion,edge label=$\nu$] (i);
					(h)--[photon,edge label=$\gamma$] (k);			
					(j)--[fermion,edge label=$\overline\nu$] (k)};
			\end{feynman}
		\end{tikzpicture}	
	}
\caption{The leading order Feynman diagrams for the photo-neutrino process.The black disk in (e and f) represents the interaction arising from effective neutrino interaction beyond  the SM.}
\label{fig0}
\end{figure}
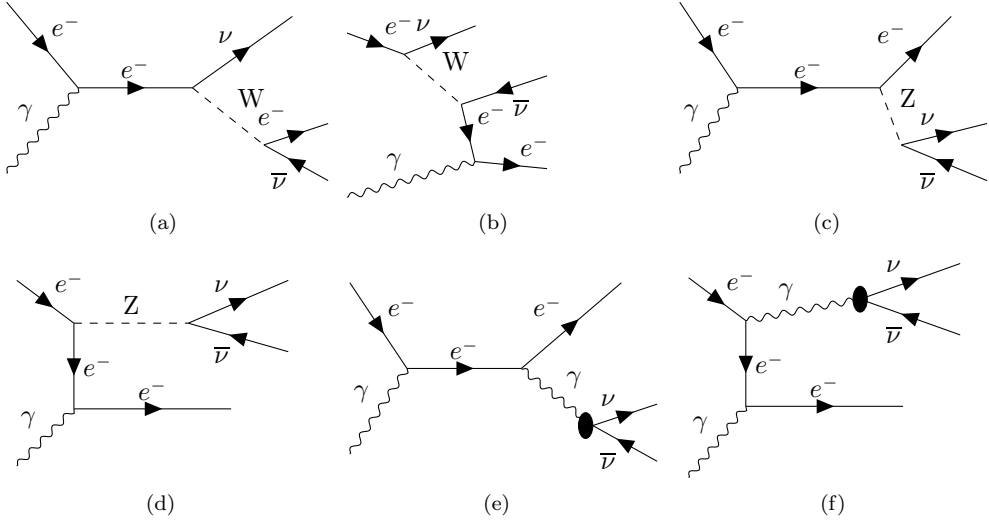

\be \baa{rcl}
\mM^{(SM)} & = & \ds - {ie G_F \over \sqrt{2}} \left[ \overline{\cu}_e(p') \gamma^\beta   (C_V-C_A \gamma_5) \frac{(\pc+\kc+m_e)}{2pk+k^2} \ec \cu_e(p) + \overline{\cu}_e (p') \ec \frac{(\pc'-\kc+m_e)}{-2p'k+k^2}  \gamma_\beta (C_V-C_A \gamma_5) \cu_e(p) \right] \\
& \times  & \ds  \overline{\cu}_\nu(q_1) \gamma_\beta  (1- \gamma_5) \cv_{\overline{\nu}}(q_2)
\eaa
\ee
where $\cu$ and $\cv$ are  Dirac spinors, $C_V={1 \over 2} +2 \sin^2 \theta_W$ and $C_A={1 \over 2}$ for $\ds \nu_e$, and $C_V=-{1 \over 2} +2 \sin^2 \theta_W$ and $C_A=-{1 \over 2}$ for $\ds \nu_\mu$ and $\ds \nu_\tau$ because the electron neutrinos interact with both $W$ and $Z$ bosons but the muon and tau neutrinos interact only with $Z$ (see the case Figure~1 (c)-(d)). In this study, we restrict ourselves by considering only the electron neutrinos. 

The matrix element  of  the (Figure~1 (e)-(f)) is given by 
\be
\mM^{(\gamma)}= \mM^{(Q)} + \mM^{(\mu)}
\ee
where
\be
\baa{rcl}
\mM^{(Q)} & = & \ds {i 4\pi\alpha \over q^2} \overline{\cu_\nu}(q_1) (\gamma^\beta - {q^\beta {\qc} \over q^2}) \left[ {q^2 \over 6} <r_\nu^2> \right] \cv_{\overline{\nu}}(q_2) \\ 
& \times & \ds (ie)^2 \left[ \overline{\cu}_e(p') \gamma^\beta    \frac{(\pc+\kc+m_e)}{2pk+k^2} \ec \cu_e(p) + \overline{\cu}_e (p') \ec \frac{(\pc'-\kc+m_e)}{-2p'k+k^2}  \gamma_\beta \cu_e(p) \right]
\eaa
\ee

\be
\baa{rcl}
\mM^{(\mu)} & = & \ds -i {2\pi\alpha \over q^2} \overline{\cu_\nu}(q_1) \sigma^{\beta \lambda} q_\lambda  \left[ \mu \right] \cv_{\overline{\nu}}(q_2) \\
& \times & \ds (ie)^2 \left[ \overline{\cu}_e(p') \gamma^\beta    \frac{(\pc+\kc+m_e)}{2pk+k^2} \ec \cu_e(p) + \overline{\cu}_e (p') \ec \frac{(\pc'-\kc+m_e)}{-2p'k+k^2}  \gamma_\beta \cu_e(p) \right]

\eaa
\ee
where $q = q_1 + q_2$, $\alpha=e^2/4\pi$ is the fine structure constant, $\ds <r_\nu^2>=<r^2>+6 \gamma_5 a$ is the mean charge radius of neutrino (in fact, it is the squared charge radius) and $\ds \mu=\mu_\nu+i \gamma_5 d_\nu$ is the neutrino magnetic moment \cite{ref14,ref17a}. 

The total matrix element of the process can be written as
\be
\mM_t = \mM^{(SM)}+ \mM^{(Q)} + \mM^{(\mu)} .
\ee
There is no interference between the helicity-conserving ($\mM^{(SM)}$ and $\mM^{(Q)}$) and helicity-flipping ($\mM^{(\mu)}$) amplitudes. Combining the helicity-conserving amplitudes and using \mbox{$\ds q_\mu J^\mu(q)=0$}, we find
\be
\baa{rcl}
\mM^{(SM)} + \mM^{(Q)} & = & \ds {ie G_F \over \sqrt{2}} \left[ \overline{\cu}_e(p') \gamma^\beta   (C'_V-C_A \gamma_5) \frac{(\pc+\kc+m_e)}{2pk+k^2} \ec \cu_e(p) + \overline{\cu}_e (p') \ec \frac{(\pc'-\kc+m_e)}{-2p'k+k^2}  \gamma_\beta (C'_V-C_A \gamma_5) \cu_e(p) \right] \\
& \times  & \ds  \overline{\cu}_\nu(q_1) \gamma_\beta  (1- \gamma_5) \cv_{\overline{\nu}}(q_2)
\eaa
\ee
where $C'_V = C_V + {\sqrt{2} \pi \alpha \over 3 G_F} <r_\nu^2>$ \cite{ref14,ref14a,ref15,ref15a,ref15b,ref15c,ref21,ref27,ref28}. We neglect the helicity-flipping amplutes because  the contribution of the magnetic moment term ($ \mM^{(\mu)}$) is very small for $\mu_{\nu_e} \le 10^{-12} \mu_B$ \cite{ref23,ref24,ref25,ref26} or $\mu_{\nu_\tau} \le 10^{-10} \mu_B$ \cite{ref26e}. 

Then the total matrix element square can be written as 
\be
|\mM_t|^2  =  |\mM^{(SM)} + \mM^{(Q)}|^2 .
\ee

The total energy loss rate ($\mQ$) or  emissivity energy  carried away by the neutrino pair per unit volume per unit time from the photo-neutrino process is given by\cite{ref3,ref4}
\be
\label{eqn:el8}
\baa{rcl}
\ds \mQ & = & \ds  {1 \over (2\pi)^{11}} \int {d^3 p \over 2E} \left[{ 2 \over \textrm{exp}[(E-\mu_e)/k_B T]+1} \right] \int {d^3 k \over 2\omega}{2 \over \textrm{exp}[\omega/k_B T]-1} \int {d^3 p' \over 2E'} \left[  1- { 1 \over \textrm{exp}[(E'-\mu_e)/k_B T]+1} \right]   \\ \\
& \times & \ds (E + \omega - E') \int  {d^3 q_1 \over 2E_\nu} \int  {d^3 q_2 \over 2E_{\overline{\nu}}} (2\pi)^4 \delta^4 (p+k-p'-q_1-q_2) {1 \over 4} \sum_{s,\epsilon} |\mM_t|^2
\eaa
\ee
where $p=(E,\overrightarrow{p})$ ,  $p'=(E',\overrightarrow{p'})$ , $k=(\omega,\overrightarrow{k})$,  $q_1=(E_\nu,\overrightarrow{q})$ and $q_2=(E_{\overline{\nu}},\overrightarrow{q'})$ are the four momentum of the incoming electron, the outgoing electron, the photon, the neutrino and  the anti-neutrino,  respectively. $\mu_e$ is the electron chemical potential, $k_B$ is the Boltzmann constant and $T$ is the Stellar temperature. The factor of $2$ in front of the electron distribution function $\ds f_e(E) = \left(\textrm{exp}[(E-\mu_e)/k_B T]+1\right)^{-1}$ and photon distribution function $f_\gamma(\omega) = \left(\textrm{exp}[\omega/k_B T]-1\right)^{-1}$ take in to account the two spin states of those particles, while the factor of ${1 \over 4}$ corresponds to the average over initial spin states. In the sum, the index $s$ indicates sums over electrons spin states, while $\epsilon$ indicates a sum over the photon spins. The factor $\left[  1- \left( \textrm{exp}[(E'-\mu_e)/k_B T]+1 \right)^{-1} \right]$ accounts for the Pauli-blocking  factor of outgoing electrons. The energy loss rate $\mQ$ in Eq. (\ref{eqn:el8})  cannot be calculated analytically for all temperature and chemical potential.

The sum of the photon polarization is carried in terms of its longitudinal and transverse components
\be
\ds \sum_{\lambda=1}^2 \epsilon^{\mu\star} \epsilon^\nu = - g^{\mu\nu} + {k^\mu k^\nu \over k^2} = P_\mL^{\mu\nu} + P_\mT^{\mu\nu} .
\ee
where longitudinal  and  transverse components are given as
\be
\ds P_\mL^{\mu\nu} =  \sum_{\lambda=1}^2 \epsilon^{\star\mu} \epsilon^\nu - P_\mT^{\mu\nu}
\ee
\be
\ds P_\mT^{\mu\nu} = \left\{ \baa{lcr} \delta^{ij} - {k^i k^j \over k^2} & \textrm{ for } & i,j=1,2,3 \\ \\ 0 & \textrm{ for } & \mu  \textrm{ or } \nu = 0 . \eaa \right. 
\ee

The photon polarization tensor mentioned above perform the following features
\bdm 
P_\mT^{\mu\beta} P_{\mL \beta \nu} = 0 \, , P_\mL^{\mu\beta} P_{\mL \beta \nu} = - P_{\mL\nu}^\mu \, , P_\mT^{\mu\beta} P_{\mT \beta \nu} = - P_{\mT \nu}^\mu \, , P_{\mL\mu}^\mu = -1 \, , P_{\mT\mu}^\mu = -2 .
\edm 
By using these relations, one can write that the longitudinal ($\mL$) and the transverse ($\mT$) component of the squared total matrix elements as:
\be
\ds \sum_{s,\epsilon} |\mM_t^{(\mL,\mT)}|^2 = 32 e^2 G_F^2 \left\{ (C_V^{'2}-C_A^2) m_e^2 |\mM_-^{(\mL,\mT)}|^2 +  (C_V^{'2}+C_A^2) |\mM_+^{(\mL,\mT)}|^2 + C_V^{'}C_A   |\mM_{+-}^{(\mL,\mT)}|^2 \right\}
\ee
here we do not present the expressions $\ds |\mM_-^{(\mL,\mT)}|^2, |\mM_+^{(\mL,\mT)}|^2$ , $\ds |\mM_{+-}^{(\mL,\mT)}|^2$  because they are very long. Our calculation is parallel to the calculation done by Dutti et al. (one can see \cite{ref11} for the details).  

Performing integration momentum $p$ and $k$, we get
\be
\ds I(p',q_1,q_2) = {1 \over \pi^2} \int {d^3 p \over 2E} \int {d^3 k \over 2\omega} f_\gamma(\omega) f_e(E) \delta^4(p+k-p'-q_1-q_2) \sum_{s,\epsilon} |\mM_t^{(\mL,\mT)}|^2 .
\ee
Then for the total energy loss we have
\be
\label{eqn14}
\ds \mQ = {1 \over (2\pi)^9} \int {d^3 q_1 \over 2E_\nu} \int {d^3 q_2 \over 2E_{\overline{\nu}}} \int {d^3 p' \over 2E'} \left[ 1 - f_e(E') \right] (E+\omega-E') I(p',q_1,q_2) .
\ee
Denote the angle between $\overrightarrow{p}+\overrightarrow{k}$ and $\overrightarrow{k}$  as $\theta_k$, we obtain
\be
\ds I(p',q_1,q_2) = {1 \over 4 (2\pi)^2} \int_0^\infty {|\overrightarrow{k}| \over \omega} d|\overrightarrow{k}| \int_0^{2\pi} d\varphi_k f_\gamma(\omega) f_e(E) {1 \over |\overrightarrow{p}+\overrightarrow{k}| } \sum_{s,\epsilon} |\mM_t^{(\mL,\mT)}|^2 .
\ee
Finally, for the differential energy loss from Eq. (\ref{eqn14}) we find
\be
\ds
{d^3 \mQ \over dE_\nu dE_{\overline{\nu}}d(\cos \theta_{\nu\overline{\nu}})} = {\pi^2 \over (2\pi)^9} E_\nu E_{\overline{\nu}} \int_0^\infty {|\overrightarrow{p'}|^2 \over E'} d |\overrightarrow{p'}| \int_{-1}^1 d(\cos \theta_e) \int_0^{2\pi} d \phi_e \,\, [1-f_e(E')](E_\nu + E_{\overline{\nu}}) I(p',q_1,q_2) .
\ee
In leading order, the dispersion relations of the photon in a plasma for the photo-neutrino process are
\bdm \ds \omega_L^2 = \omega_P^2 \, , \quad \omega_T^2 = \omega_P^2 + |\overrightarrow{k}|^2
\edm
where $\omega_P$ is the plasma frequency.

It is helpful to describe the foremost physical scales in the photo-neutrino production process of the neutrino pairs under limited circumstance for obtaining a qualitative and in many cases a quantitative understanding of the neutrino emissivity as a function of the temperature. We can procure the degenerate plasma which occurs in all temperature levels at sufficiently high densities, and a non-degenerate relativistic plasma that has high temperature but low density.

We will investigate the term $\mQ_\mT$ for transverse case; the analysis for the longitudinal case can be performed along the similar way. Although we cannot calculate the total energy loss rate analytically for all $T$ and $\mu_e$, we will consider for evaluating ($\mQ$) in three different regions where it can be done analytically,namely, for nondegenerate, intermediate and degenerate electrons.

\subsection{The Nondegenerate Case}

This case occurs at enough low densities for which $\mu_e - m_e \ll T$. For $T \ge 10^{10} K$, the electron mass can be neglected in comparison to $\mu_e$. In the relativistic case, the electron density and the plasma frequency are given as \cite{ref29}
\bdm 
\ds n_e = {\mu_e \over 3 \pi^2}(\mu_e^2 + T^2 \pi^2) \simeq {\mu_e T^2 \over 3}
\edm 
\bdm
\ds \omega_p^2 = {4 \alpha \over 3 \pi }(\mu_e^2 + {T^2 \pi^2 \over 3})  \simeq  4 \pi \alpha {T^2 \over 9} \, .
\edm

In this case both electron momentum and energy are of order of $E \simeq |\overrightarrow{p}| \simeq T$ and similarly both the photon momentum and energy 
are $\omega \simeq |\overrightarrow{k}| \simeq T$. So, we obtain for the squared total matrix element as
\be
\ds \sum_{s,\epsilon} |\mM_t^\mT|^2 \simeq 256 \pi \alpha G_F^2  (C_V^{'2} + C_A^2) E  E_\nu E_{\overline{\nu}} / E' .
\ee

Then performing the phase space integration we have
\be
\ds \mQ_\mT = {20 \alpha G_F^2 (C_V^{'2} + C_A^2) \over 3 (2\pi)^6} T^9 \int_0^\infty {x^2 dx \over e^x+1} \int_0^\infty {y dy \over e^y-1} \int_0^{x+y} {(x+y-z)^3 dz \over e^{-z}+1}
\ee
where $x \equiv E/T, y \equiv \omega/T$ and $z \equiv E'/T$. Analytical approximation to these integrals can be obtained  by replacing  $\ds {1 \over e^{-z}+1 } \rightarrow 1$. In this limit we find
\bdm
\int_0^\infty {x^2 dx \over e^x+1} \int_0^\infty {y dy \over e^y-1} \int_0^{x+y} {(x+y-z)^3 dz \over e^{-z}+1} = {63 \over 256} \Gamma(7)\zeta(7)\Gamma(2)\zeta(2) + {37 \over 22} \Gamma(6)\zeta(6)\Gamma(3)\zeta(3)+{73 \over 32} \Gamma(5)\zeta(5)\Gamma(4)\zeta(4) 
\edm
where $\Gamma(n)$ and $\zeta(n)$ are standard gamma and zeta functions, respectively.

In this limit, we obtain
\be
\ds \mQ_\mT \simeq {20042 \over 3 (2\pi)^6} \alpha G_F^2 (C_V^{'2} + C_A^2) T^9 .
\ee

\subsection{The Intermediate Case}
In this case $\mu_e > T > \omega_p$, the energy loss rate $\mQ_\mT$ is strongly dependent on density . In this density range, the dominant energies are $E \simeq |\overrightarrow{p}| \simeq \mu_e , \omega \simeq |\overrightarrow{k}| \simeq T , E_\nu \simeq |\overrightarrow{q}| \simeq T$. With this energy scales the squared total matrix element is estimated as
\be
\ds \sum_{s,\epsilon} |\mM_t^\mT|^2 \simeq 256 \pi \alpha G_F^2  (C_V^{'2} + C_A^2) {1 \over 4 \mu_e T \omega_p^2} T^4 \mu_e^2 .
\ee
Similar to the nondegenerate case, the phase space integration brings
\be
\ds \mQ_\mT \simeq {32 \alpha G_F^2 (C_V^{'2} + C_A^2) \over 3 (2\pi)^6} {T^9 \mu_e^2 \over \omega_p^2} \zeta(5) \Gamma(5,\omega_p/T) 
\ee
where $\Gamma(n,m)$ is the incomplete gamma function.

In the case of the maximum emissivity, the ratio ${\omega_p \over T} \ll 1$, for energy loss rate we get 
\be
\ds \mQ_\mT \simeq {768 \over 3 (2\pi)^6} \alpha G_F^2 (C_V^{'2} + C_A^2) {T^9 \mu_e^2 \over \omega_p^2} e^{-\omega_p/T} .
\ee

\subsection{The Degenerate Case}

As the last case, $\mu_e$ and $\omega_p \gg T$ and $m_e$ where $\ds \mu_e \simeq (3 \pi^2 n_e)^{1/3}$, $\omega_p \simeq \sqrt{4 \alpha/3\pi} \mu_e \simeq \mu_e/18$, one can obtain \cite{ref29}
\bdm \ds n_e = {1 \over \pi^2} \int_0^\infty dp p^2 \left[\left(\textrm{exp}({E-\mu_e \over k_B T})+1\right)^{-1} - \left(\textrm{exp}({E'-\mu_e \over k_B T})+1\right)^{-1}\right]. \edm The Pauli blocking factor of outgoing electron provides that the electrons lie close to the Fermi surface. In this case electrons are elastically scattered exchanging only the 3-momentum with the photon and outgoing neutrinos. Under this circumstances, it is  expected that both  electron energies would be $E \simeq E' \simeq |\overrightarrow{p}| \simeq |\overrightarrow{p'}| \simeq \mu_e$. As a results, the photon's entire energy is converted to the neutrino-antineutrino pair energy. Hence the photon energy and momentum are $\omega \simeq \omega_p \simeq E_\nu + E_{\overline{\nu}}$ and $|\overrightarrow{k}| \simeq T, E_\nu \simeq |\overrightarrow{q}| \simeq \omega_p /2$. In this limit the squared matrix element can be approximated by
\be
\ds \sum_{s,\epsilon} |\mM_t^\mT|^2 \simeq 16 \pi \alpha G_F^2  (C_V^{'2} + C_A^2) \omega_p^2 .
\ee
Then  energy loss rate becomes
\be
\ds \mQ_\mT \simeq {4 \alpha G_F^2 (C_V^{'2} + C_A^2) \over 3 (2\pi)^6} \omega_p^6 T^3 e^{-\omega_p/T}.
\ee

\section{Numerical Results and Discussion}

In this section, we will present our numerical results for the energy loss rate for considered three different cases in terms of tables and figures. For the numerical calculations we have selected the range for the neutrino charge radius $<r_\nu^2>$ as $[0 , 100] \cdot 10^{-32} cm^2$ and for the temperature $T$ as $[10^{8} K, 10^{11}K]$ (see \cite{ref15,ref15a,ref15b,ref15c}). 

\begin{table}[!h]
	\centering
	\begin{tabular}{|c|l|l|l|l|l|}
		\hline
		\multicolumn{2}{|c|}{} &\multicolumn{4}{|c|}{$<r_\nu^2> \cdot 10^{-32} cm^2$} \\
		\cline{3-6}
		\multicolumn{2}{|c|}{}  & 0 & 2.5 & 10 & 100 \\\hline
		\multirow{4}{*}{\rotatebox{90}{$T$}}  & $10^8 K$ & $1.17908 \times 10^{4}$   & $1.29761 \times 10^{4}$ & $1.69596 \times 10^{4}$   & $1.14784 \times 10^{5}$  \\\cline{2-6}    
		& $10^9 K $ &  $1.17908 \times 10^{13}$  & $1.29761 \times 10^{13}$ & $1.69596 \times 10^{13}$  & $1.14784 \times 10^{14}$  \\\cline{2-6}
		& $10^{10} K $ &  $1.17908 \times 10^{22}$ & $1.29761 \times 10^{22}$ &  $1.69596 \times 10^{22}$  & $1.14784 \times 10^{23}$  \\\cline{2-6}
		& $10^{11} K$ & $1.17908 \times 10^{31}$ & $1.29761 \times 10^{31}$  &  $1.69596 \times 10^{31}$  & $1.14784 \times 10^{32}$\\
		\hline  
	\end{tabular}
	\caption{Energy loss rate values ($\mQ_\mT (erg/cm^3 \cdot s)$) for the nondegenerate case.}
	\label{tbl1}
\end{table}

\begin{table}[!h]
	\centering
	\begin{tabular}{|c|l|l|l|l|l|}
		\hline
		\multicolumn{2}{|c|}{} &\multicolumn{4}{|c|}{$<r_\nu^2> \cdot 10^{-32} cm^2$} \\
		\cline{3-6}
		\multicolumn{2}{|c|}{}  & 0 & 2.5 & 10 & 100 \\\hline
		\multirow{4}{*}{\rotatebox{90}{$T$}}  & $10^8 K$ & $1.00253 \times 10^{5}$   & $1.10332 \times 10^{5}$ & $1.44202 \times 10^{5}$   & $9.75972 \times 10^{5}$  \\\cline{2-6}    
		& $10^9 K $ &  $1.00253 \times 10^{14}$  & $1.10332 \times 10^{14}$ & $1.44202 \times 10^{14}$  & $9.75972 \times 10^{14}$  \\\cline{2-6}
		& $10^{10} K $ &  $1.00253 \times 10^{23}$ & $1.10332 \times 10^{23}$ &  $1.44202 \times 10^{23}$  & $9.75972 \times 10^{23}$  \\\cline{2-6}
		& $10^{11} K$ & $1.00253 \times 10^{32}$ & $1.10332 \times 10^{32}$  &  $1.44202 \times 10^{32}$  & $9.75972 \times 10^{32}$\\
		\hline  
	\end{tabular}
	\caption{Energy loss rate values ($\mQ_\mT (erg/cm^3 \cdot s)$) for the intermediate  case.}
	\label{tbl2}
\end{table}

\begin{table}[!h]
	\centering
	\begin{tabular}{|c|r|l|l|l|l|}
		\hline
		\multicolumn{2}{|c|}{} &\multicolumn{4}{|c|}{$<r_\nu^2> \cdot 10^{-32} cm^2$} \\
		\cline{3-6}
		\multicolumn{2}{|c|}{}  & 0 & 2.5 & 10 & 100 \\\hline
		\multirow{4}{*}{\rotatebox{90}{$T$}}  & $10^{10} K$ & $2.60048 \times 10^{20}$   & $2.8619 \times 10^{20}$ & $3.74047 \times 10^{20}$   & $2.53158 \times 10^{20}$  \\\cline{2-6}    
		& $3 \times 10^{10} K $ &  $5.11852 \times 10^{24}$  & $5.63309 \times 10^{24}$ & $7.36237 \times 10^{24}$  & $4.98292 \times 10^{25}$  \\\cline{2-6}
		& $5 \times 10^{10} K $ &  $5.07906 \times 10^{26}$ & $5.58966 \times 10^{26}$ &  $7.30561 \times 10^{26}$  & $4.9445 \times 10^{27}$  \\\cline{2-6}
		& $10 \times 10^{10} K$ & $2.60048 \times 10^{29}$ & $2.8619 \times 10^{29}$  &  $3.74047 \times 10^{29}$  & $2.53158 \times 10^{32}$\\
		\hline  
	\end{tabular}
	\caption{Energy loss rate values ($\mQ_\mT (erg/cm^3 \cdot s)$) for the degenerate  case.}
	\label{tbl3}
\end{table}

Firstly, in Table~1, Table~2 and Table~3, we displayed the energy loss rate values ($\mQ_\mT (erg/cm^3 \cdot s)$) for all three cases for the different values of $T$ and $<r_\nu^2> \cdot 10^{-32} cm^2$. It is seen that the charge radius contribution is considerable at high  temperature in all considered cases. Also, one can observe that there is an effective change in the energy loss rate for the degenerate case compared to the other one.

\begin{figure}[!h]
	\centering
	\subfigure[$T=10^8K$]{\includegraphics[width=5.3cm]{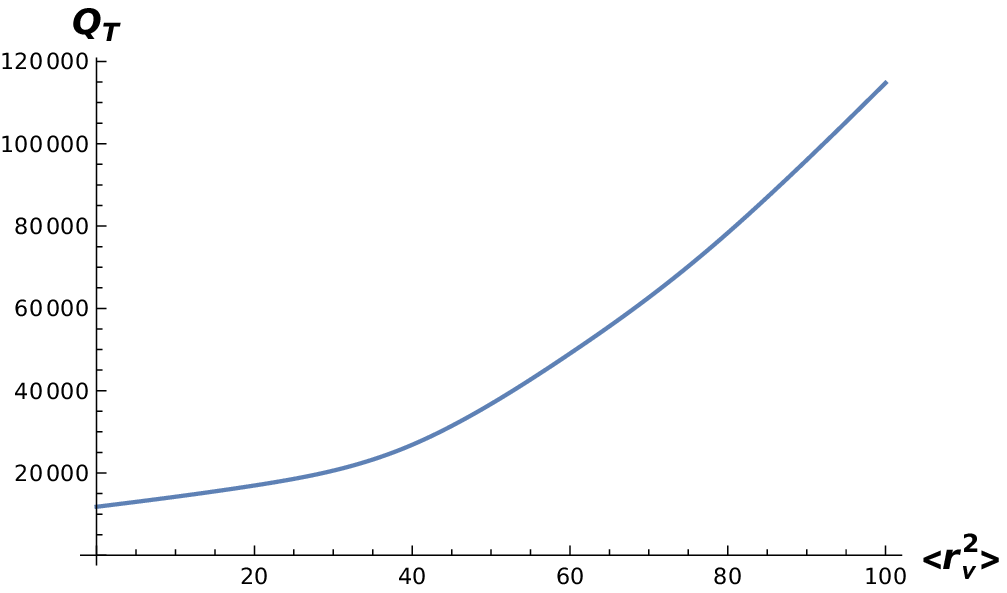}}
	\subfigure[$T=0.5 \cdot 10^9K$]{\includegraphics[width=5.3cm]{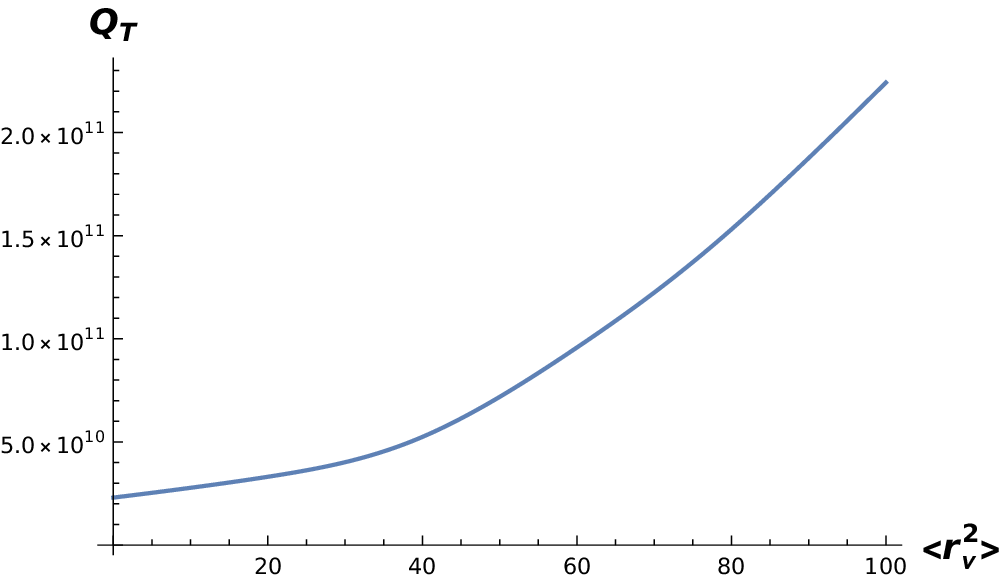}}
	\subfigure[$T=10^9K$]{\includegraphics[width=5.3cm]{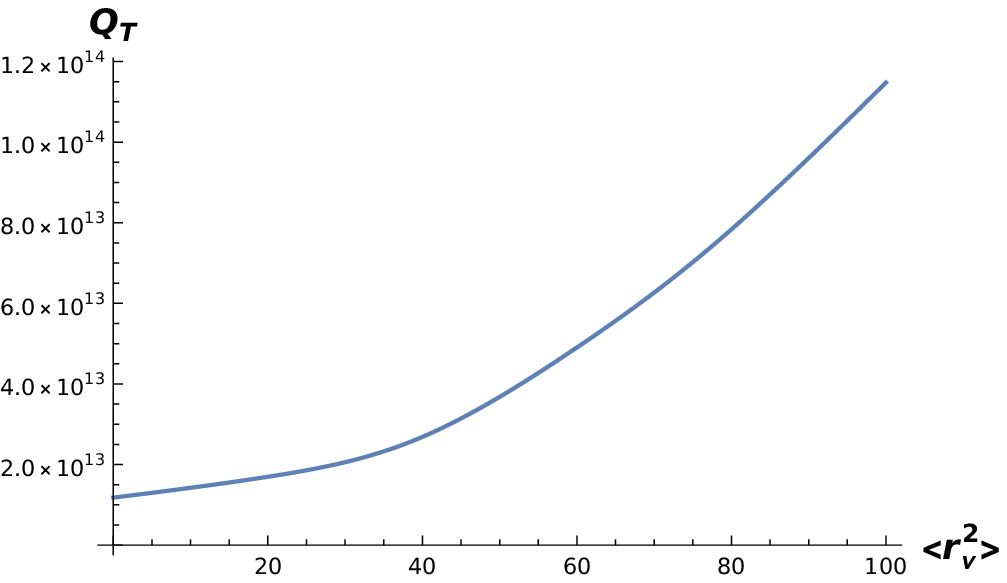}}
\caption{Energy loss rate values($\mQ_\mT (erg/cm^3 \cdot s)$) versus charge radius ($<r_\nu^2> \times (10^{-32}cm^2$) for different values of the temperature ($T$) for the nondegenerate case.}
	\label{fig01}
\end{figure}

\begin{figure}[!h]
	\centering
	\subfigure[$<r_\nu^2>=0$]{\includegraphics[width=5.3cm]{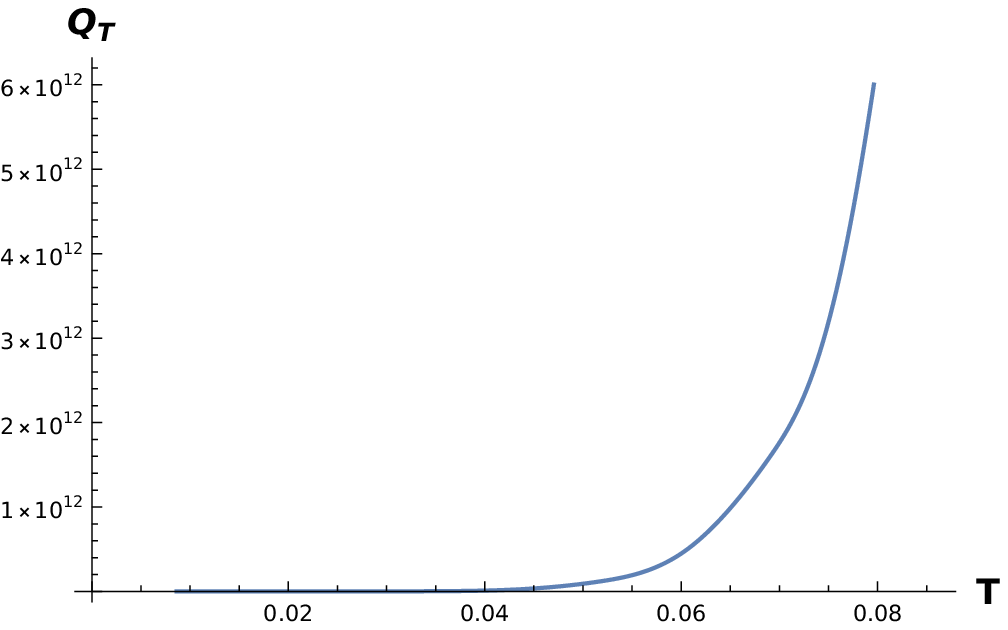}}
	\subfigure[$<r_\nu^2> \times (10^{-31}cm^2 )$]{\includegraphics[width=5.3cm]{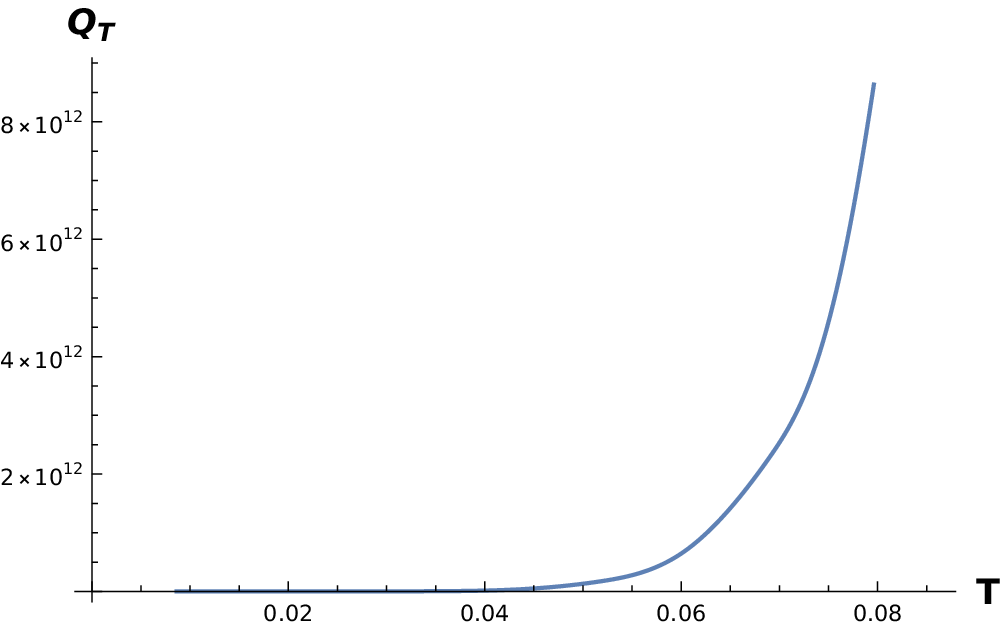}}
	\subfigure[$<r_\nu^2> \times (10^{-30}cm^2 )$]{\includegraphics[width=5.3cm]{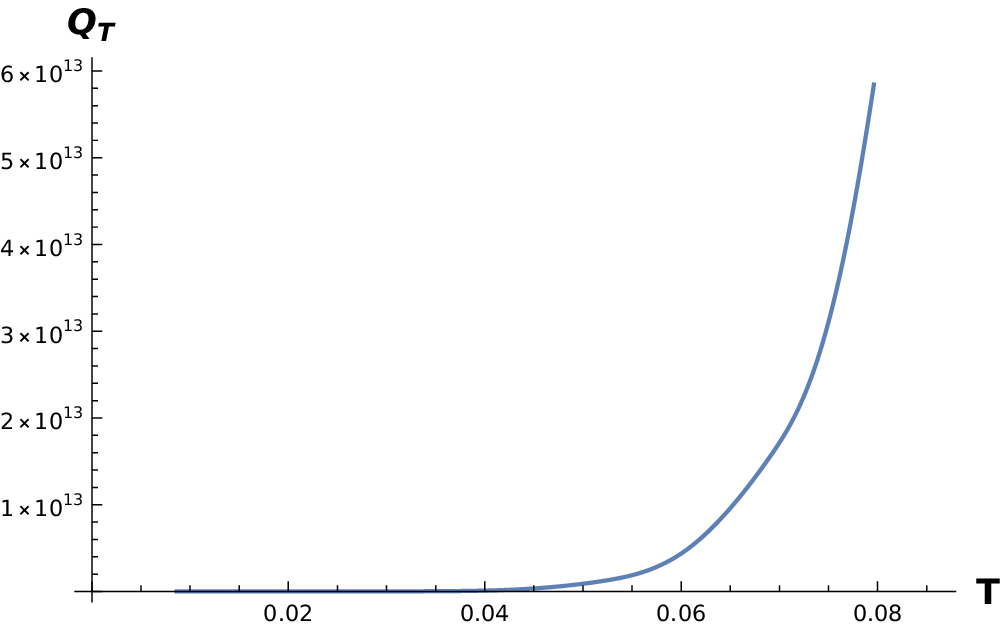}}
\caption{Energy loss rate values($\mQ_\mT (erg/cm^3 \cdot s)$) versus temperature ($T(10^9 K)$) for different values of the charge radius ($<r_\nu^2> $) for the nondegenerate case.}
	\label{fig02}
\end{figure}

In Figure~2  and Figure~3 we visualized the effect of $T$ and $<r_\nu^2> \times (10^{-32}cm^2$ on the energy loss rate values($\mQ_\mT (erg/cm^3 \cdot s)$). As expected from the obtained formulas, the change with respect to $T$ is more stronger compared to change with respect to $<r_\nu^2>$.

\begin{figure}[!h]
	\centering
	\subfigure[nondegenerate case]{\includegraphics[width=5cm]{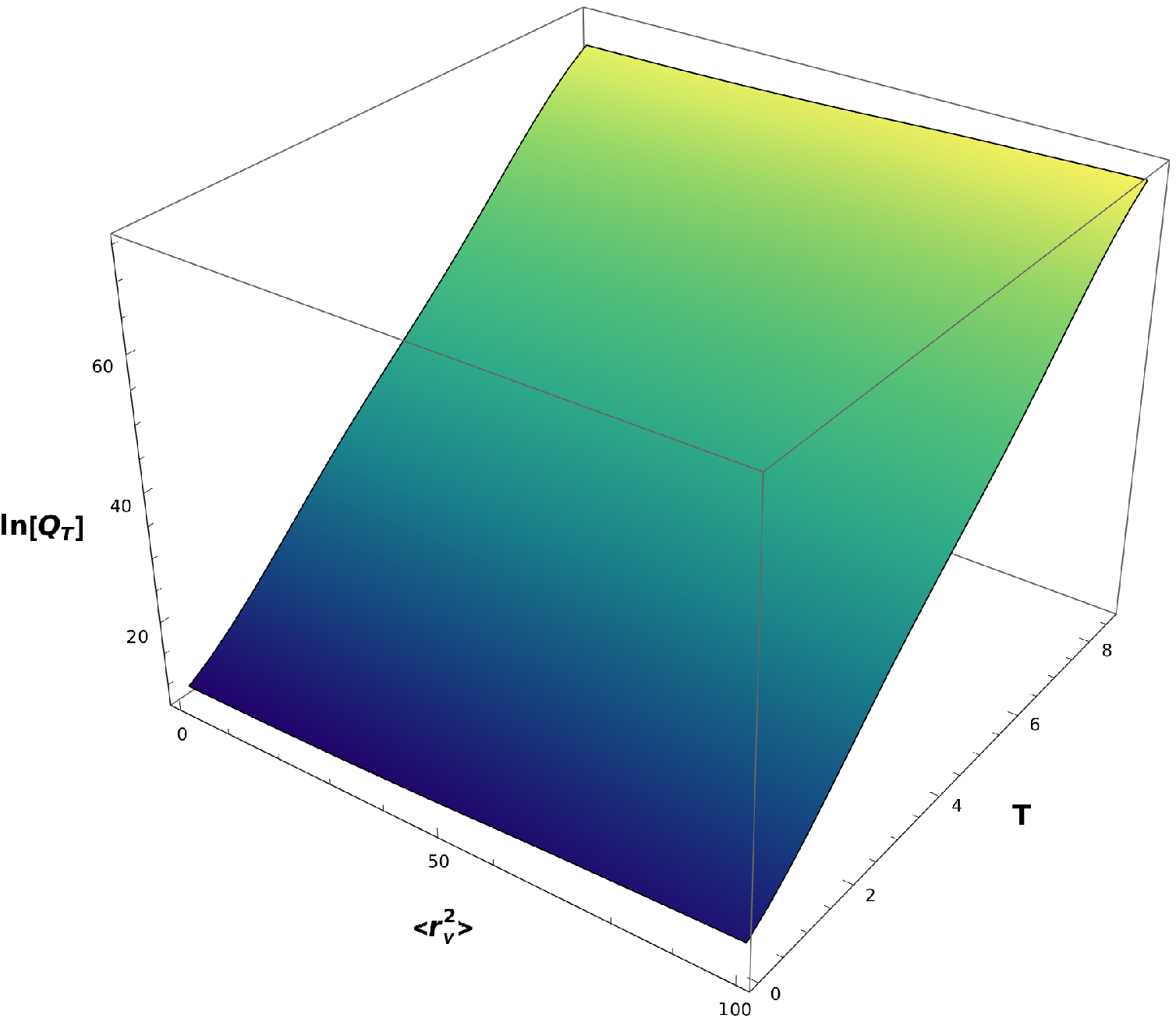}}
	\subfigure[intermediate case]{\includegraphics[width=5cm]{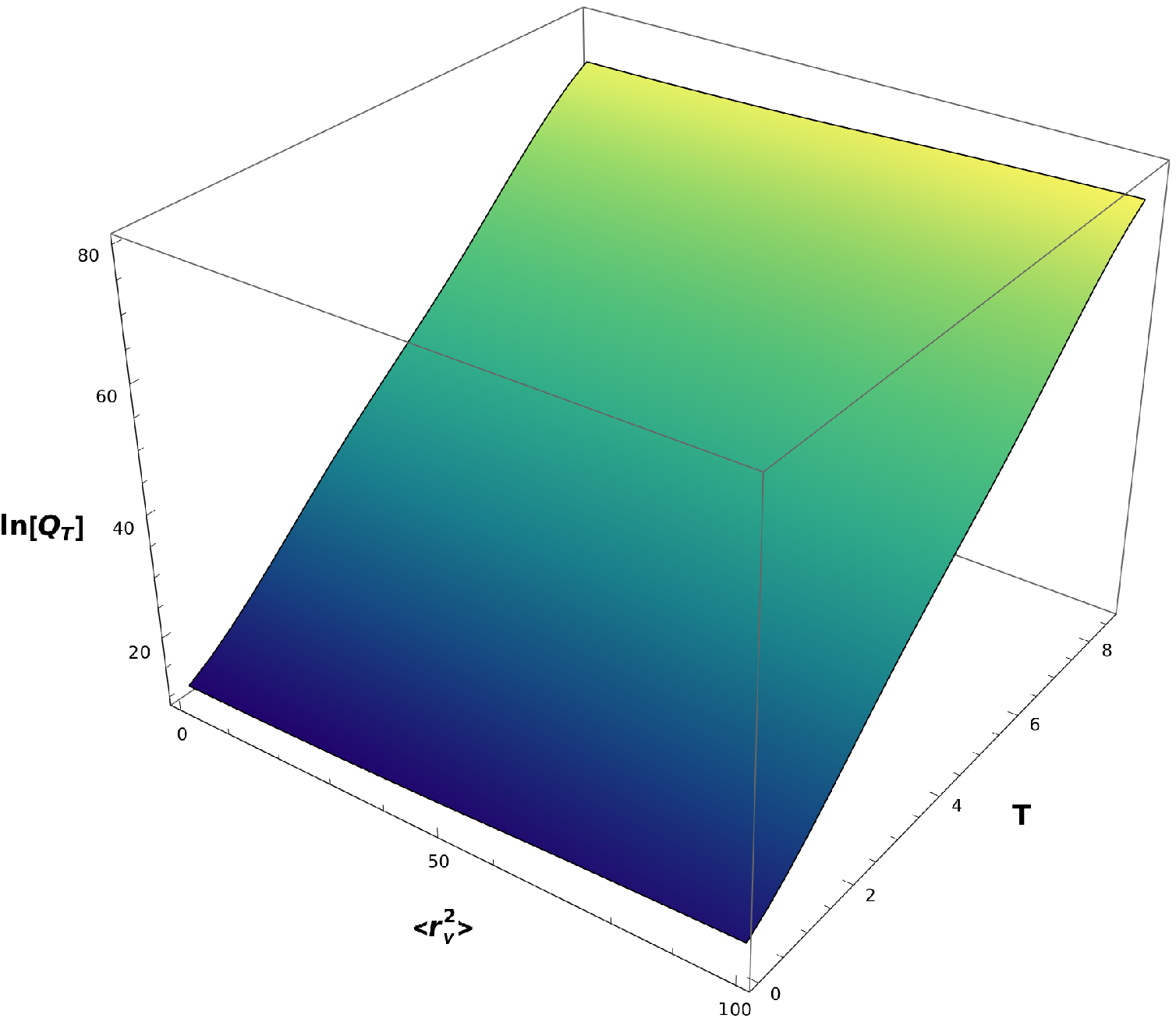}}
	\subfigure[degenerate case]{\includegraphics[width=5cm]{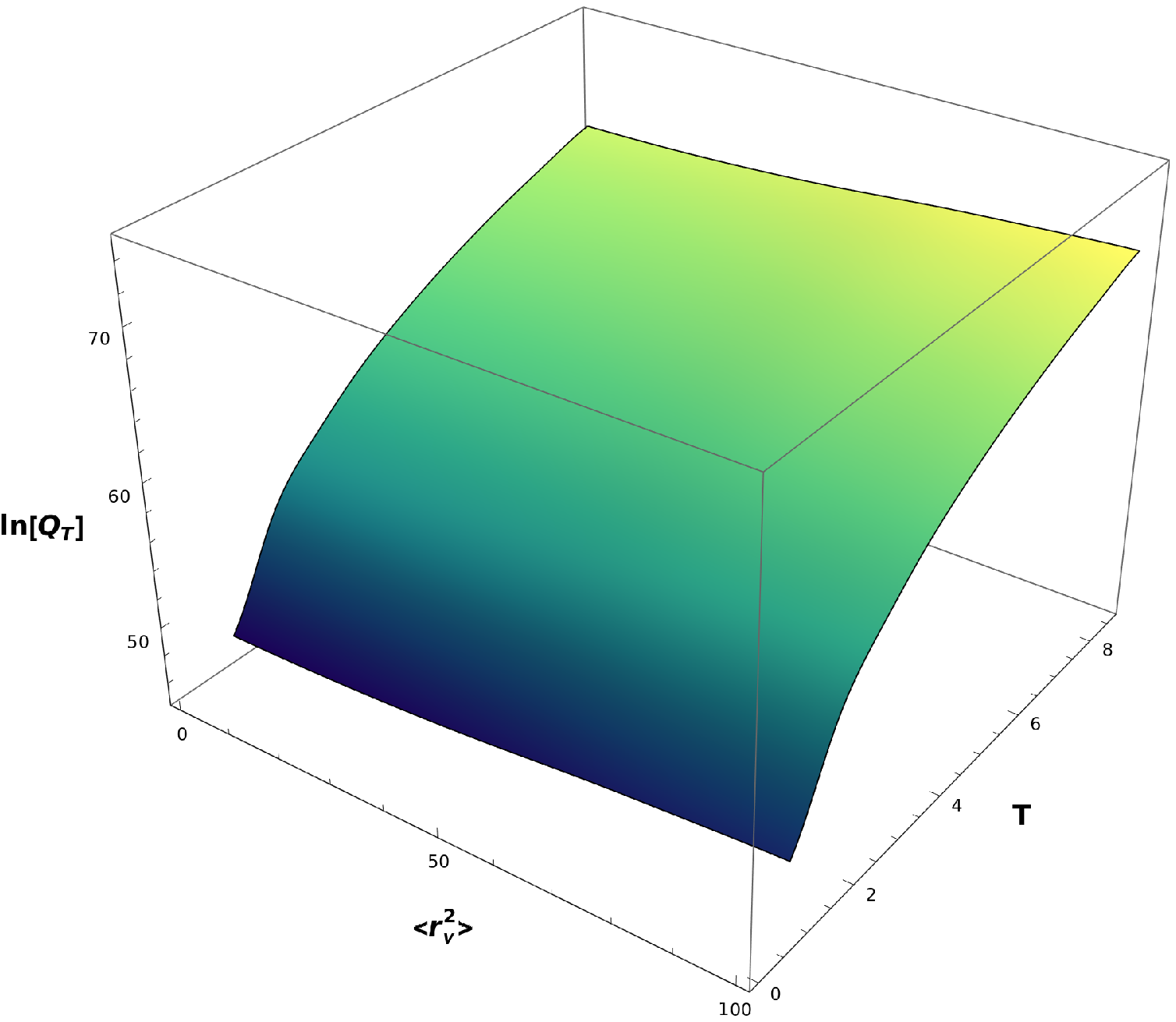}}
\caption{Logarithm of the total energy loss rate values($\ln(\mQ_\mT (erg/cm^3 \cdot s))$) versus charge radius and temperature ($T(10^9 K)$) for the  $<r_\nu^2> \times (10^{-32}cm^2)$ in (a) and  $<r_\nu^2> \times (10^{-30}cm^2)$ in (b) and (c).}
	\label{fig03}
\end{figure}

Finally, in Figures (\ref{fig03}) we plotted the 3-D graph of $\ln(\mQ_\mT)$ for the values of temperature and charge radius. Similar to the table results, it is seen that the change in the the total energy loss rate is different in the degenerate case compared to the two other cases due to the division energy of the photon by the neutrino and anti-neutrino.

\section{Conclusion}

As a conclusion, in this study we have calculated energy loss rate of photo-neutrino process, considering neutrino charge  radius (or anapole moment) effects. It is obtained that, the contribution for the charge radius change the result about $\sim 10\%$. The neutrinos' magnetic moment contribution is very small. We observed that if the neutrino magnetic moment  have $\mu_\nu = 10^{-6} \mu_B$ order, then the effect of the indicated term becomes dominant and the contribution should be taken under consideration. However, this value is still bigger than the tau-neutrino's magnetic moment.

\clearpage

\end{document}